\documentclass[vecphys]{svmult}

\usepackage{graphicx}        
\usepackage[bottom]{footmisc}

\begin{document}

\title*{A principal component analysis approach  to the morphology of Planetary Nebulae}
\titlerunning{PCA approach to the morphology of PNe}
\author{Stavros Akras\inst{1,2}\and
Panayotis Boumis\inst{1}}
\institute{Institute of Astronomy \& Astrophysics, National Observatory of Athens, Athens, Greece
\texttt{arkas@astro.noa.gr},
\texttt{ptb@astro.noa.gr}
\and Department of Physics, University of Patras, Rio-Patras, Greece}
%
%
\maketitle

\begin{abstract}
Principal Component Analysis (PCA) is a well--known technique used to
decorrelate a set of vectors. It has been applied to explore the star
formation history of galaxies or to determine distances of
mass--lossing stars. Here we apply PCA to the optical data of
Planetary Nebulae (PNe) with the aim of extracting information about
their morphological differences. Preliminary analysis of a sample of
55 PNe with known abundances and morphology shows that the second
component (PC2), which results from a relation produced by the
parameters log(N/O), initial and final mass of PNe, is depending on
the morphology of PNe. It has been found that when log(N/O) $< -0.18$
the PNe's nitrogen is low independently on the oxygen abundance for
either Bipolar ({\it B}), Elliptical ({\it E}) or Round ({\it R})
PNe. An interesting result is that both {\it E} and {\it R} PNe have
log(N/O) $< 0$ while only {\it B} PNe show negative and positive
values. Consequently, {\it B} PNe are expected to have higher nitrogen
values than the {\it E} and {\it R} PNe. Following that and a second
sample of 35 PNe, n$_{\rm e}$ is also found to be higher in {\it B}
PNe. Also, in all PNe morphologies PC2 appears to have a minimum at
0.89 and PNe's initial mass at 2.6 M$_{\odot}$. 5--D diagrams between
PCAs components and physical parameters are also presented. More
results will follow while simple models will be applied in order to
try to give a physical meaning to the components. 
\end{abstract}

\section{Introduction}
\label{sec:1}
Planetary Nebulae (PNe) are powerful tools in the study of the
evolutionary scenario of intermediate mass stars. They play an
important role in the chemical enrichment history of the interstellar
medium and many efforts have been devoted to determine the physical
parameters of Galactic PNe (like T$_{\rm e}$, n$_{\rm e}$, T$^{\star}$,
L$^{\star}$, distance, abundances; \cite{journal1} and references
therein). For this study a number of methods have been used either by
using the observational results directly or by developing simulation
models like CLOUDY \cite{journal2}. Using the latter, the values of
the physical parameters can be determined (i.e. \cite{journal3}) but
not any possible correlation among them. So far, only statistical
methods had been used in order to find correlations between the
parameters. However, there is a well--known technique (PCA) which can
be used in order to study the possible correlation between these
parameters. PCA technique uses a sample of observed parameters and
creates a new sample of independent components which are linear
combination from the previous parameters.

\section{PCA methods and 5--D diagrams}
\label{sec:2}
PCA is a useful tool in statistical studies, especially when there are
many observed parameters. It has been applied to explore the stars
formation history or to describe and classify the stellar spectra. The
basic idea of this method is that it minimizes a set of observed
parameters to a new set of independent parameters. The number of these
new components reveals the true dimension of the space generated by
the observed parameters. Also, a 5--D diagram has been developed in
order to study if there is any correlation between the new components
and the measured parameters. These diagrams are like common 3--D
diagrams with the difference that, they illustrate two more variables
using different colours and sizes.

\section{Preliminary results}
\label{sec:3}
In this work, we applied PCA method to a set of optical data of 55 PNe
with known abundances and morphology \cite{journal4}, in order to
extract information correlated to their morphological differences. The
new principal components PC$_{1}$, PC$_{2}$ and PC$_{3}$ ((1)-(3))
resulted by using a set of parameters (log(N/O), core Mass M$_{\rm c}$ and
initial Mass M$_{\rm i}$) which are correlated according to the star
formation history (\cite{journal5} \& \cite{journal6}). It can be
seen from (1)--(3) that the difference between PC$_{1}$, PC$_{2}$ and
PC$_{3}$ is that each one is strongly depended on one of these
parameters only.

\begin{equation}
{\rm PC_1=0.89 \times log(N/O)+0.43 \times M_c - 0.12\times M_i}
\end{equation}

\begin{equation}
{\rm PC_2=-0.43 \times log(N/O)+0.90 \times M_c +0.07\times M_i}
\end{equation}

\begin{equation}
{\rm PC_3=0.14 \times log(N/O) -0.07 \times M_c + 0.99\times M_i}
\end{equation}

The PC$_{1}$ and PC$_{3}$ components do not provide any significant
result, while PC$_{2}$ shows that it depends on the morphology of
PNe. From (2), we produced the diagrams shown in Figures 1(a),(b) \&
2, where it can be seen that when log(N/O) $< -$0.18, the PNe's
nitrogen abundance is low independently on the oxygen for either {\it
B}, {\it E} or {\it R} morphology. Moreover, the log(N/O) is negative,
if the PN morphological type is either {\it E} or {\it R}, while it
takes both negative and positive values in the case of {\it B}
morphology. PC$_{2}$ component appears to have a minimum value at 0.89
for all morphological types, resulting to an initial mass for the
progenitor star of 2.6 M$_{\odot}$. PCA method was also applied to
another sample of 35 PNe \cite{journal7}, using a set of different
parameters (abundance $\epsilon$(N), $\epsilon$(S) and electron
density n$_{\rm e}$) which are important to PNe's study. In this case,
the new principal components PC$_{4}$, PC$_{5}$ and PC$_{6}$
((4)--(6)) were calculated to:

\begin{equation}
{\rm PC_4=0.91 \times \epsilon(N) - 0.41 \times \epsilon(S) - 0.05 \times n_e}
\end{equation}

\begin{equation}
{\rm PC_5=0.41 \times \epsilon(N) - 0.91 \times \epsilon(S) + 0.04 \times n_e}
\end{equation}

\begin{equation}
{\rm PC_6=0.06 \times \epsilon(N) + 0.02 \times \epsilon(S) + 0.99 \times n_e}
\end{equation}

In Figure 3, we present a 5--D diagram with PC$_{4}$, PC$_{5}$ and
PC$_{6}$ on X,Y and Z axis respectively and $\epsilon$(S), n$_{\rm e}$
on different colours and sizes. In particular, it can be seen that
$\epsilon$(S) increases when PC$_{5}$ increases (the circular colour
bar on the right of Figure 3 shows the increase of $\epsilon$(S) as we
go from $0^{\rm o} $ ($\epsilon({\rm S})=$4.2) to $360^{\rm o}$
($\epsilon({\rm S})=$7.7)). Also, it is clear that n$_{\rm e}$ increases
(larger size means higher value of n$_{\rm e}$) linearly with PC$_{4}$
and PC$_{5}$. This 5--D representation shows that there is a strong
dependence between PC$_{4}$ and PC$_{5}$, $\epsilon({\rm S})$ and
n$_{\rm e}$ since for the same value of $\epsilon({\rm S})$ (same
colour i.e. blue), we result to a linear equation ${\rm PC_5=0.8\times
PC_4 + 6.04}$ (7). By using (4),(5) and (7) we produce the
relationship ${\rm n_e= 4\times \epsilon(N)+75.5- 15.5
\times\epsilon(S)}$ (8). Since in that case $\epsilon({\rm S})$ is
constant, then n$_{\rm e}$ increases when $\epsilon({\rm N})$
increases. Following this result and the one from the previous sample,
it appears that the {\it B} PNe must have higher n$_{\rm e}$ values
than either {\it E} or {\it R} PNe. These preliminary results show
that the use of new components which are correlated with physical
parameters can provide useful information therefore must be explored
further in the future.

\section{Acknowledgments}
\label{sec:4}
The authors would like to thank I. Ouranos for providing the 5--D
model (animation). SA acknowledgments funding by the European Union
and the Greek Ministry of Development in the framework of the
programme 'Promotion of Excellence in Research Institutes ($2^{\rm nd}$
Part)'.

\begin{figure}
\centering
\includegraphics[height=12.5cm]{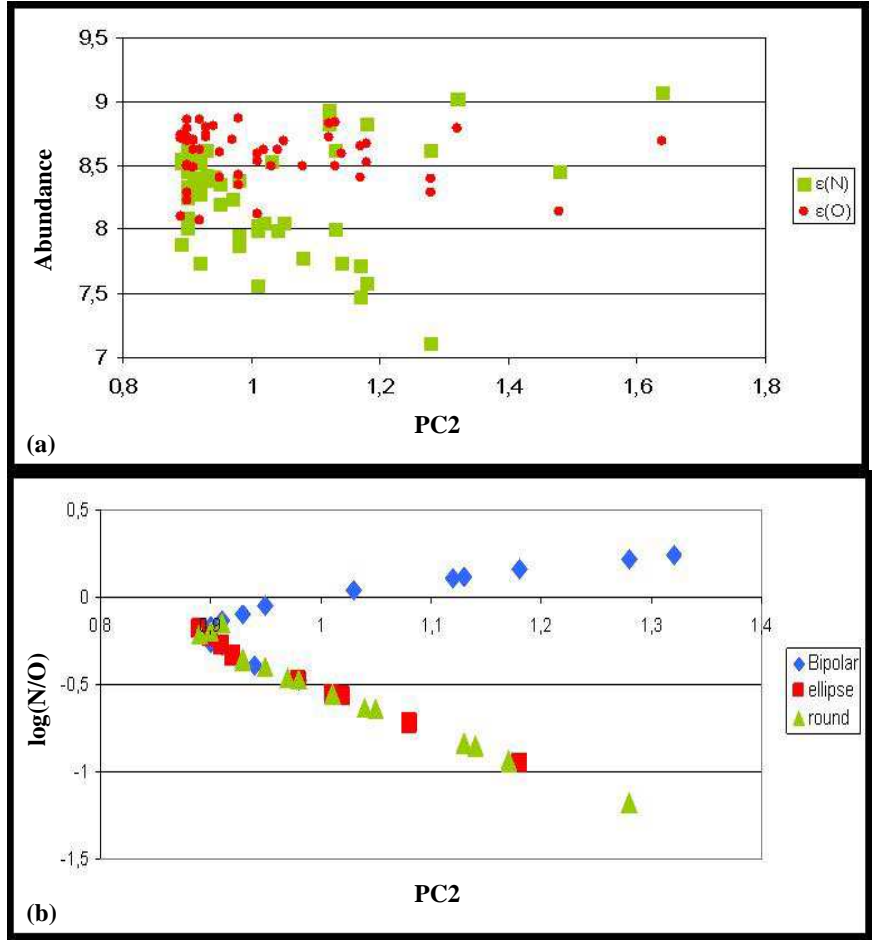}
\caption{(a) Diagrams of Nitrogen and Oxygen abundances vs
PC$_{2}$~component for the 55 PNe. (b) Diagram of log(N/O) vs PC$_{2}$
where the different morphological types are shown (see text).}
\label{fig:1}       
\end{figure}

\begin{figure}
\centering
\includegraphics[height=6cm]{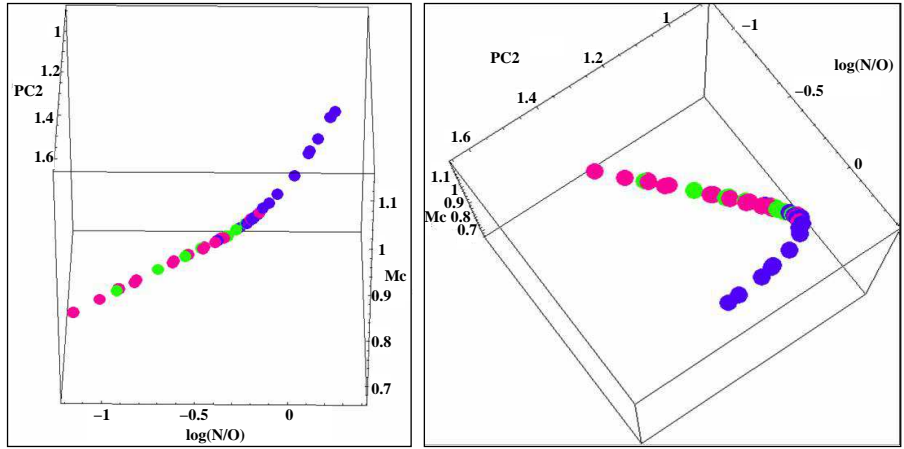}
\caption{Two different 4D representations of Fig. 1(b).}
\label{fig:2}   
\end{figure}

\begin{figure}
\centering
\includegraphics[height=10cm]{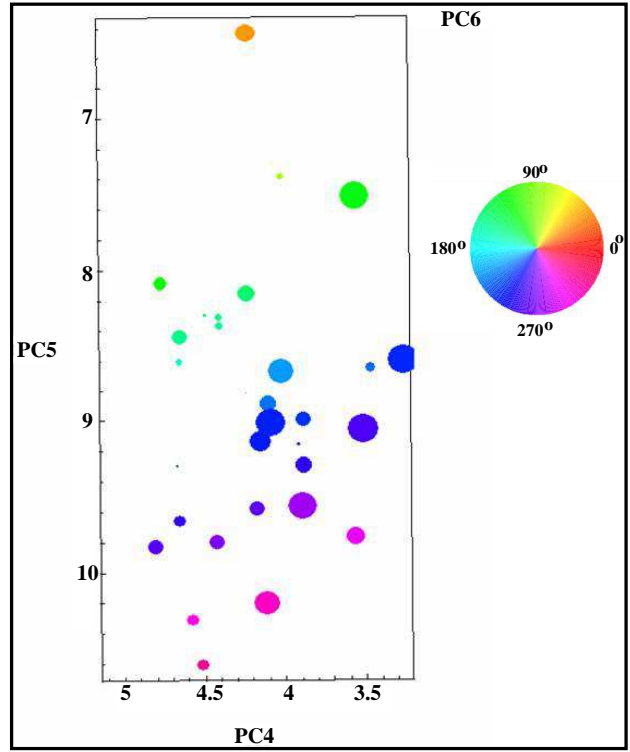}
\caption{PC$_{4}$~vs PC$_{5}$; $\epsilon$(S) and n$_{\rm e}$~are shown
with different colours and sizes (see text).}
\label{fig:3}       
\end{figure}



\end{document}